\documentclass[times]{stvrauth}

\usepackage{moreverb}
\usepackage[colorlinks,bookmarksopen,bookmarksnumbered,citecolor=red,urlcolor=red]{hyperref}
\usepackage{subcaption}
\usepackage{cite}
\usepackage{tabularx}

\usepackage{paralist}
\setdefaultenum{(i)}{}{}{}

\usepackage{listings}
\usepackage{xspace}

\newcommand{\juge}{\textsc{JUGE}\xspace}
\newcommand{\runtool}{\texttt{runtool}\xspace}
\newcommand{\randoop}{\textsc{Randoop}\xspace}
\newcommand{\mytool}{\textsc{MyTool}\xspace}
\newcommand{\evosuite}{\textsc{EvoSuite}\xspace}
\newcommand{\evosuitedse}{\evosuite (dynamic symbolic execution)\xspace}
\newcommand{\bbc}{\textsc{BBC}\xspace}
\newcommand{\mosa}{\textsc{MOSA}\xspace}
\newcommand{\dynamosa}{\textsc{DynaMOSA}\xspace}
\newcommand{\fuzzbench}{\textsc{FuzzBench}\xspace}
\newcommand{\grt}{\textsc{GRT}\xspace}
\newcommand{\sushi}{\textsc{Sushi}\xspace}
\newcommand{\tardis}{\textsc{Tardis}\xspace}
\newcommand{\utbot}{\textsc{UtBot}\xspace}
\newcommand{\kex}{\textsc{Kex}\xspace}
\newcommand{\jtexpert}{\textsc{JTExpert}\xspace}
\newcommand{\tthree}{\textsc{T3}\xspace}
\newcommand{\ttwo}{\textsc{T2}\xspace}
\newcommand{\pitest}{\textsc{PITest}\xspace}
\newcommand{\jacoco}{\textsc{JaCoCo}\xspace}
\newcommand{\defectsforj}{\textsc{Defects4J}\xspace}

\newcommand{\eg}{\textit{e.g.,~}}
\newcommand{\Eg}{\textit{E.g.,~}}
\newcommand{\ie}{\textit{i.e.,~}}

\newcommand{\etal}{\textit{et al.}\xspace}
\newcommand{\etc}{\textit{etc.}\xspace}
\newcommand{\wrt}{\textit{w.r.t.~}}

\newcommand\BibTeX{{\rmfamily B\kern-.05em \textsc{i\kern-.025em b}\kern-.08em
T\kern-.1667em\lower.7ex\hbox{E}\kern-.125emX}}

\def\volumeyear{2022}

\begin{document}

%\runningheads{A.~N.~Other}{A demonstration of the \journalabb\ class file}

\title{\juge: An Infrastructure for Benchmarking\\Java Unit Test Generators}

\author{Xavier Devroey\affil{1}\affil{6}\corrauth, % ORCID: 0000-0002-0831-7606
Alessio Gambi\affil{2}\affil{8}, % ORCID: 0000-0001-6539-2162
Juan Pablo Galeotti\affil{3}, % ORCID: 0000-0002-0747-8205
Ren{\'e} Just\affil{4}, % ORCID: 0000-0002-5982-275X
Fitsum Kifetew\affil{5}, % ORCID: 0000-0003-1860-8666
Annibale Panichella\affil{6}, % ORCID: 0000-0002-7395-3588
and Sebastiano Panichella\affil{7} % ORCID: 0000-0003-4120-626X
}

\address{%
\affil{1} NADI, University of Namur, rue de Bruxelles 61, 5000 Namur, Belgium
\break
\affil{2} University of Passau, Innstrasse 41, 94032 Passau, Germany
\break
\affil{3} Universidad de Buenos Aires, Argentina
\break
\affil{4} University of Washington, Seattle, USA
\break
\affil{5} Fondazione Bruno Kessler, via Sommarive 18, 38123 Trento, Italy
\break
\affil{6} Delft University of Technology, Postbus 5, 2600 AA Delft, The Netherlands
\break
\affil{7} Zurich University of Applied Sciences, Steinberggasse 13, 8400 Winterthur, Switzerland
\break
\affil{8} IMC University of Applied Sciences, Am Campus Krems, Trakt G, 3500 Krems an der Donau, Austria
}

\corraddr{Xavier Devroey, NADI, University of Namur, rue de Bruxelles 61, 5000 Namur, Belgium. E-mail: xavier.devroey@unamur.be}

% Doesn't work with multiple sponsors -_-
% \cgsn{NWO Vici project ``TestShift''}{VI.C.182.032}
% \cgsn{EU Horizon 2020 ICT-10-2016-RIA ``STAMP''}{731529}
% \cgsn{EU Horizon 2020 ``COSMOS''}{957254}

\begin{abstract}
    Researchers and practitioners have designed and implemented various automated test case generators to support effective software testing. Such generators exist for various languages (\eg Java, C\#, or Python) and various platforms (\eg desktop, web, or mobile applications). The generators exhibit varying effectiveness and efficiency, depending on the testing goals they aim to satisfy (\eg unit-testing of libraries vs. system-testing of entire applications) and the underlying techniques they implement. 
In this context, practitioners need to be able to compare different generators to identify the most suited one for their requirements, while researchers seek to identify future research directions. This can be achieved by systematically executing large-scale evaluations of different generators. However, executing such empirical evaluations is not trivial and requires substantial effort to select appropriate benchmarks, setup the evaluation infrastructure, and collect and analyze the results. 
In this Software Note, we present our \emph{JUnit Generation Benchmarking Infrastructure} (\juge) supporting generators (search-based, random-based, symbolic execution, \etc) seeking to automate the production of unit tests for various purposes (validation, regression testing, fault localization, \etc). The primary goal is to reduce the overall benchmarking effort, ease the comparison of several generators, and enhance the knowledge transfer between academia and industry by standardizing the evaluation and comparison process.
Since 2013, several editions of a unit testing tool competition, co-located with the Search-Based Software Testing Workshop, have taken place where \juge was used and evolved. As a result, an increasing amount of tools (over ten) from academia and industry have been evaluated on \juge, matured over the years, and allowed the identification of future research directions. Based on the experience gained from the competitions, we discuss the expected impact of \juge in improving the knowledge transfer on tools and approaches for test generation between academia and industry. Indeed, the \juge infrastructure demonstrated an implementation design that is flexible enough to enable the integration of additional unit test generation tools, which is practical for developers and allows researchers to experiment with new and advanced unit testing tools and approaches.

\end{abstract}

\keywords{\juge; unit test generation; evaluation infrastructure; benchmarking}

\maketitle

%%%%%%%%%%%%%%%%%%%%%%%%%%%%%%%%%%%%%%%%%%%%%%%%%%%%%%%%%%%%%%%%%%%%%%%%%%%%%

%%%%%%%%%%%%%%%%%%%%%%%%%%%%%%%%%%
\section{Introduction}
\label{sec:introduction}
%%%%%%%%%%%%%%%%%%%%%%%%%%%%%%%%%%

Over the last decades, researchers have come up with various techniques to automate the generation of test cases. In particular, unit test generators seek to automate the production of tests for various purposes (validation, regression testing, fault localization, \etc) using different techniques, including random search (\eg~\cite{Pacheco2007,Ma2016c}), search-based software testing (\eg~\cite{Fraser2011,Alshahwan2018,derakhshanfar2020botsing}), and symbolic (\eg~\cite{papadakis2010automatic,Braione2017}) and concolic execution (\eg~\cite{yun2018qsym,sen2007concolic}).

Juristo \etal \cite{Juristo2004} identified three essential features each empirical evaluation should contribute to the software testing empirical body of knowledge. First, the evaluation should be fully defined, and the data should be analyzed with appropriate techniques to interpret the results. Second, the programs used for the evaluation and the setup and variables considered should represent the reality of the practice. Third, an evaluation should be replicable and come with a replication package to confirm previous results and reach an acceptable level of confidence in the hypothesis.

Similarly, to bridge the gap with industry, automated test case generators must come with solid evidence that the approach can also be applied in practice. For instance, evidence-based software engineering \cite{Dyba2005} can help practitioners make informed decisions about the choice of a generator based on the current best evidence from research. Those pieces of evidence come from empirical evaluations identifying the strengths and weaknesses of various generators.

In the case of unit test generators, conducting an empirical evaluation is not trivial. It requires an extensive manual effort to collect benchmarks (\ie Java classes for which to generate test cases), setup the evaluation and the evaluation infrastructure, collect and analyze the produced unit tests, and compare the results against the state-of-the-art. The primary goal of our JUnit Generation Benchmarking Infrastructure (\juge) \cite{xavier_devroey_2021_4904393} is to reduce the overall effort and ease the comparison of several generators by standardizing the evaluation process. This standardization allows researchers to meet the requirements, enabling an effective contribution to the empirical body of knowledge in software testing.

\juge is suited for evaluating and comparing \emph{fully automated} black, white, and gray-box unit test generators. For instance, in past editions of the tool competition, \juge has been applied to evaluate various types of tools relying on a variety of approaches, including search-based \cite{Fraser2011, Panichella2018, Sakti2015}, random-based \cite{Ma2016c, Pacheco2007, Prasetya2014}, and symbolic execution \cite{Braione2017, Braione2018}. In a nutshell, the generator takes the source code or the binaries of a Java project as input and generates unit tests for a given class or set of classes. A time budget limits the generation, and the generated tests are compared \wrt their structural coverage and mutation score. The benchmarks, the tests, and the intermediate results can be saved and archived to be added to a replication package and enable future comparisons without re-executing all the generators, thanks to the standardized evaluation process implemented in \juge.

\juge has been initially developed in the context of the tool competition, co-located with the Search-Based Software Testing (SBST) workshop. It has been built to be \emph{extensible} and configurable to be used with different test case generators and deployed in different environments. It relies on standardized processes, including an \emph{adapter} mechanism to run a tool and process the generated tests. \emph{Isolation} of the generator and test executions are handled through containerization (using Docker). Similarly, \emph{scalability} of the evaluations relies on the parallelization of different containers, which allows relying on standard technologies (\eg Docker commands and dashboards) to handle the overall evaluation smoothly. 

Since 2013, ten editions of the tool competition have taken place and used the \juge infrastructure to evaluate and compare automated unit test generators \cite{Competition2013, Competition2014, Competition2015, Competition2016, Competition2017, Competition2018, Competition2019, Competition2020, Competition2021, Competition2022}. Consequently, \juge has been improved and evolved over the years to integrate the latest advances from academia to enhance the comparison and best practices from industry to achieve high automation. 
Several tools have entered the competition \cite{Tool2013EvoSuite, Tool2013T3, Tool2014EvoSuite, Tool2014T3, Tool2015EvoSuite, Tool2015GRT, Tool2015JTExpert, Tool2015MOSA, Tool2015T3, Tool2016EvoSuite, Tool2016JTExpert, Tool2016T3, Tool2017EvoSuite, Tool2017JTExpert,  Tool2018EvoSuite, Tool2018T3, Tool2019EvoSuite, Tool2019SUSHI, Tool2019T3, Tool2020EvoSuite, Tool2021EvoSuite, Tool2021EvoSuiteDSE, Tool2021Kex, Tool2021UtBot, Tool2022BBC, Tool2022EvoSuite, Tool2022Kex, Tool2022UtBot} and matured over the years by fixing bugs evidenced by the evaluations using the \juge infrastructure, but also by confronting the various approaches to different benchmarks to discover areas for improvement and future research directions. The current implementation is openly available on GitHub\footnote{\url{https://github.com/JUnitContest/JUGE}} and on Zenodo for long-term storage \cite{xavier_devroey_2021_4904393}.

The remainder of this paper is structured as follows: Section \ref{sec:background} discusses the background and related work of empirical evaluation and comparison of test case generators. Section \ref{sec:infrastructure} presents the challenges and requirements for building \juge, as well as the design and implementation choices made to address those challenges. Section \ref{sec:evaluation} provides guidelines for setting up an evaluation with \juge (examples of evaluations setups using \juge are discussed in Appendix \ref{sec:example}). Section \ref{sec:impact} presents the impact of \juge so far, while Section \ref{sec:discussion} discusses the lessons learned from building \juge and running evaluation with it, as well as the selection of suitable benchmarks (\ie classes under test), and the future work. Finally, Section \ref{sec:conclusion} concludes the paper. 

%%%%%%%%%%%%%%%%%%%%%%%%%%%%%%%%%%%%%%%
\section{Background and related work}
\label{sec:background}
%%%%%%%%%%%%%%%%%%%%%%%%%%%%%%%%%%%%%%%

When designing a new test case generation technique, conducting empirical evaluations is paramount to position this new technique in the current software testing body of knowledge \cite{Juristo2004, Vos2012}. When the technique gains in maturity, developers will also rely on those empirical evaluations to make informed decisions about choosing a tool relevant to their industrial context \cite{Dyba2005, Alshahwan2018}. For instance, Melo \etal \cite{Melo2019} designed a recommender for concurrent software testing techniques based on the characteristics of the software under test and the current body of knowledge in concurrent software testing.

% -------------------------------------------
\subsection{Empirical evaluation guidelines}
% -------------------------------------------

Over the years, several guidelines, benchmarks, and infrastructures have been developed to design, execute, and assess test case generators. For instance, Arcuri and Briand \cite{Arcuri2014} defined guidelines for using statistical tools when evaluating and comparing randomized algorithms, which is the case with many automated test case generators. 
In their systematic review of the empirical evaluation of search-based test case generation, Ali \etal \cite{Ali2010a} identify the elements that should be reported in study designs. They found that search-based software testing has been focused on structural coverage and unit testing and that empirical studies should adopt a more rigorous and standardized execution and reporting approach. In particular, studies should account for random variation in the results using appropriate statistical hypothesis testing. They should then compare the techniques with other baselines to conclude that it brings any advantage. 

More recently, in a significant effort to improve the review process in software engineering, Ralph \etal \cite{Ralph2020a} defined \textit{Empirical Standards} listing specific attributes expected when conducting empirical evaluation following a given research methodology. The empirical evaluation of automated test cases generation is classified under the umbrella of \textit{Optimization Studies in Software Engineering}: \ie \textit{research studies that focus on the formulation of software engineering problems as search problems and apply optimization techniques to solve such problems} \cite{Ralph2020a}. Among the essential characteristics of such studies, the standards require comparing the approach to an appropriate baseline and the distribution of the dataset (\ie the benchmarks used for the evaluation, if possible, and the results).

\juge contributes to the general effort of improving the quality and reproducibility of empirical evaluations for unit test generators by 
(i) standardizing the evaluation process, using appropriate data analysis techniques, 
and (ii) enabling easy distribution of the benchmarks (\ie classes under tests used for the evaluation) and results, including the test cases, coverage, and mutation analysis, and statistical analysis for future comparisons and reproductions. Section \ref{sec:evaluation} discusses the guidelines to design, execute, and report the results of an empirical evaluation using our infrastructure.

Similar efforts have been pursued in other areas of automated testing. Recently, \fuzzbench \cite{Metzman2021}, an open fuzzer benchmarking platform-as-a-service, has received much attention after its deployment, thanks to the support of Google. Similarly to \juge, \fuzzbench provides a standard procedure to benchmark fuzzer, using coverage and fault coverage on representative program datasets. It has been designed to be scalable and ensure fair resource allocation and platform independence. \juge achieves the same goals by relying on containerization via Docker. It allows running the different generators in isolation while relying on standard technologies and tools (\eg Docker dashboards) to manage scalability, resource allocations, and independence from the underlying platform (more details about the infrastructure will be provided in Section \ref{sec:infrastructure}). 

% -----------------------------------------------
\subsection{Comparison of test case generators}
% -----------------------------------------------

Besides structural coverage, like lines or branch coverage, empirical evaluations rely on mutation analysis to compare different test case generators \cite{Andrews2006}. Mutation analysis \cite{DeMillo1978} applies mutation operators, \eg replacing an arithmetic operator, to a program under test to produce faulty variants (\ie mutants) and executes a test suite on those variants. If a test fails on a particular mutant, this mutant is considered as \textit{killed}. The mutation score, \ie the ratio of killed mutants to the total number of mutants, is used to measure the faults detection capabilities of the test suite \cite{Just2014}. 

For now, \juge supports structural coverage and mutation analysis of the generated tests. Other kinds of automated analysis can be plugged into the infrastructure's extensible architecture. Additionally, all the generated test suites are saved using a unique identifier and can be collected for additional manual inspection. 

% --------------------------------------------
\subsection{Benchmarks for software testing}
% --------------------------------------------

Empirical evaluations can be performed on various benchmarks (\ie classes under test). For instance, Fraser and Arcuri built \textsc{SF110} \cite{Fraser2014b}, a corpus of 23,886 classes from 110 open-source projects used to evaluate and compare unit test generators. Other benchmarks follow a different approach by using actual bugs extracted from Java software systems. For instance, \defectsforj \cite{Just2014b} is a collection of reproducible bugs and a supporting infrastructure widely used for evaluating software testing and debugging approaches. In its latest version (v2.0.0), \defectsforj contains 835 bugs from 17 Java software systems \cite{Gay2020}.
Similarly, \textsc{BugSwarm} \cite{Dmeiri2019} is a toolkit designed to mine reproducible failures and corresponding fixes to evaluate fault-detection, localization, and repair approaches.

\juge supports the definition of customized benchmarks. For instance, previous editions of the tool competition have used classes from \defectsforj's projects and classes collected from open-source projects. Section \ref{subsec:benchmarkselection} discusses guidelines to select classes under test for an empirical evaluation based on our experience in the tool competition and related work.

%%%%%%%%%%%%%%%%%%%%%%%%%%%%%%%%%%
\section{JUGE Infrastructure}
\label{sec:infrastructure}
%%%%%%%%%%%%%%%%%%%%%%%%%%%%%%%%%%

We describe hereafter the challenges and requirements for building an automated infrastructure like \juge. Based on those requirements, we present the architecture and implementation of \juge.

%-------------------------------------
\subsection{Challenges and requirements}
\label{subsec:requirements}
%-------------------------------------

Building an evaluation infrastructure like \juge poses several challenges. Indeed, to be reusable and have an impact on automated unit test generation, such infrastructure has to meet several requirements. We present and discuss the most important ones in the following paragraphs. 

\newcommand{\challenegeextensible}{(\textbf{C1})\xspace}
\paragraph{Extensible and configurable \challenegeextensible.}
\juge has to be extensible to adapt to different research contexts. For instance, \juge has been used for the unit testing competition (described in Section \ref{sec:impact}) for which the overall execution time is limited. It should also be practical for large-scale evaluations running over several weeks. 
For this, adding new generators to the infrastructure should be easy to compare them to the state-of-the-art. Similarly, it should be easy to configure \juge to use a given set of classes under test (\ie benchmarks) for an evaluation. 
Finally, it should be easy to extend \juge to use different measures on the generated tests and add new ones to adapt to the different research questions and hypotheses driving the empirical evaluation. 

\newcommand{\challenegeisolation}{(\textbf{C2})\xspace}
\paragraph{Isolation \challenegeisolation.}
The execution of automated test case generators may have side effects on the rest of a system \cite{Fraser2013}. For instance, a generator using all available resources by default can negatively impact other generators running in parallel. Additionally, unexpected behavior, such as failures caused by faults in the generator, can cause damage to the system. Moreover, evaluating the generated tests requires executing them, which might also cause undesirable side effects. \juge has to include mechanisms to isolate the execution of the generator and the generated test from the rest of the host system to avoid any undesirable side effects or damages.  

\newcommand{\challenegeperformance}{(\textbf{C3})\xspace}
\paragraph{Performance and scalability \challenegeperformance.}
\juge should be able to run on various platforms depending on the resources available. For instance, it should run on a standalone machine but also on a computationally intensive server or in a distributed setting on several servers in parallel. It should also scale to extensive empirical evaluations involving several tools and hundreds of classes under test. 

\newcommand{\challenegestandard}{(\textbf{C4})\xspace}
\paragraph{Standardization \challenegestandard.}
Finally, evaluations relying on \juge should be standardized to ease replication. This includes configuration of \juge for a given evaluation, recording, analysis, and reporting of the data, and archiving to provide a companion artifact for the evaluation.

%-------------------------------------
\subsection{Implementation of JUGE}
\label{subsec:implementation}
%-------------------------------------

Given the identified requirements, we developed \juge. \juge is \textit{extensible} and can be \textit{configured} for evaluating and comparing \emph{fully automated} black, white, and grey-box unit test generators \challenegeextensible. The generator expects as input the source code or the binaries of a Java project and generates unit tests for a given class or set of classes. The generation is limited by a time budget provided as input to the generator. The overall execution is limited by a global timeout (\ie to twice the time budget) to take the pre and post-processing of the generator into account. For each benchmark (\ie class under test), \juge runs the test case generator with the given time budget. \juge can be configured to repeat the executions a given amount of times to balance \textit{performance} and \textit{scalability} \challenegeperformance, depending on the available resources and the required statistical power of the results. Once the generation is completed, \juge can measure structural coverage, perform mutation analysis of the generated tests, and compare the different generators using sound statistical analysis in a \textit{standard} manner \challenegestandard. Moreover, the benchmarks, generated tests, and other data can be collected and stored in an artifact. 

\juge is open-source, available on GitHub\footnote{\url{https://github.com/JUnitContest/JUGE}} and packaged as a Docker image to ensure \textit{isolation} from the host system \challenegeisolation. It contains scripts and tools supporting 
\begin{inparaenum}
\item the \emph{generation of unit tests} for a given set of classes under test and time budget;
\item the \emph{coverage and mutation analysis} of the generated tests;
and \item the \emph{statistical analysis} and comparison of different unit test generators. 
\end{inparaenum}

\begin{figure}[t]
    \centering
    \includegraphics[width=0.75\textwidth]{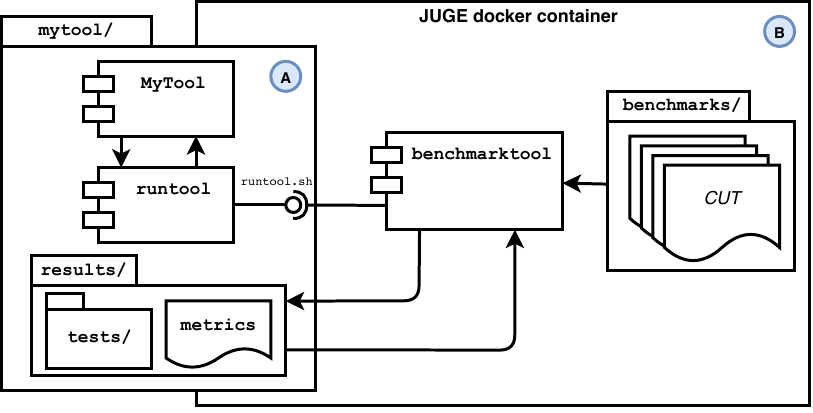}
    \caption{\juge architecture overview}
    \label{fig:architecture}
\end{figure}

As illustrated in Figure \ref{fig:architecture}, \juge relies on an \emph{adapter}, called \runtool, to wrap specific calls to a unit test generator (\mytool in \autoref{fig:architecture}). This adapter offers an interface to the \texttt{benchmarktool}, in charge of orchestrating the evaluation of the unit test generator. 
The communication between the host and the \texttt{\juge docker container} (\textit{B} in \autoref{fig:architecture}) is done via a common folder (\textit{A} in \autoref{fig:architecture}), mounted in the file tree structure of the image. This folder contains the executable binaries of the unit test generator and its \runtool adapter. 
The generated tests, the metrics, and the statistical analysis results are saved in a subfolder (\texttt{results/}) to be made available to the host. The classes under test and the corresponding configuration file are saved in the Docker container (\texttt{benchmarks/}). Hence, to evaluate multiple tools, one can reuse the same container and only has to mount different folders, each containing the unit test generator and its \runtool adapter.

%-------------------------------------
\subsection{Unit test generation}
\label{subsec:unittestgeneration}
%-------------------------------------

One of the main challenges when building \juge was to define a generic protocol for the generation of unit tests able to handle various unit test generators. We rely on a set of conventions and a generic communication protocol between the \texttt{benchmarktool} and the \runtool adapter. 

\paragraph{Conventions.}
By convention, the common folder (\textit{A} in \autoref{fig:architecture}) has to be named after the generator (\texttt{mytool/} in our example) and mounted in the \texttt{/home/} directory of the Docker container. For any generator, unit tests have to be generated in \texttt{/home/mytool/temp/testcases/}. Unit tests have to be stored as one or more Java test files containing JUnit tests for each class under test. Each Java test file has to declare a public class with a zero-argument public constructor, annotate test methods with \texttt{@Test}, and declare test methods public. Additional files may be saved to \texttt{/home/mytool/temp/data/} for later offline analysis (\eg for debugging of the generator). 

\begin{lstlisting}[float=t,
    numbers=none,
    label=lst:ITestingTool,
    caption={\texttt{ITestingTool} adapter interface},
    language=Java]
public interface ITestingTool {

    /**
     * List of additional class path entries required by a testing tool.
     * @return List of directories/jar files.
     */
    public List<File> getExtraClassPath();

    /**
     * Initialize the testing tool, with details about the code to be tested (SUT).
     * Called only once.
     * @param src       Directory containing source files of the SUT.
     * @param bin       Directory containing class files of the SUT.
     * @param classPath List of directories/jar files (dependencies of the SUT).
     */
    public void initialize(File src, File bin, List<File> classPath);

    /**
     * Run the test tool, and let it generate test cases for a given class.
     * @param cName      Name of the class for which unit tests should be generated.
     * @param timeBudget How long the tool must run to test the class (in miliseconds).
     */
    public void run(String cName, long timeBudget);

}
\end{lstlisting}

\paragraph{Adapter.}
To make the liaison between the \texttt{benchmarktool} and the \runtool adapter, the folder \texttt{/home/mytool/} must contain a \runtool executable script or binary (\eg \texttt{runtool.sh} in Figure \ref{fig:architecture}) that will be called by the \texttt{benchmarktool} to start the generation of unit tests. Typically, \texttt{runtool.sh} contains a single command launching the adapter. A customizable implementation of \textit{runtool} is provided in the source code repository of our infrastructure.\footnote{ \url{https://github.com/JUnitContest/JUGE/tree/master/runtool}} 
The create an adapter for a new tool, one has to implement the different methods of the \texttt{ITestingTool.java} interface described in Listing \ref{lst:ITestingTool}.\footnote{\url{https://github.com/JUnitContest/JUGE/blob/master/runtool/src/main/java/sbst/runtool/ITestingTool.java}}  
This effort should be minimal for any reasonably well-implemented tool. For instance, the implementation for \randoop is 102 lines long, against around 150 lines of code for  \evosuite-based implementations. 
The methods are then called, as specified by the protocol described in the following paragraph.

% \begin{lstlisting}[float=t,
%     numbers=none,
%     label=lst:runtoolmytool,
%     caption={\mytool \runtool script}]
% #!/bin/bash

% APACHE_EXECS_LIB=lib/org/apache/commons/commons-exec/1.2/commons-exec-1.2.jar
% TOOL=lib/runtool-1.0.0-SNAPSHOT.jar
% java -cp $TOOL:$APACHE_EXECS_LIB  sbst.runtool.Main
% \end{lstlisting}

\begin{figure}[t]
    \centering
    \includegraphics[width=120mm]{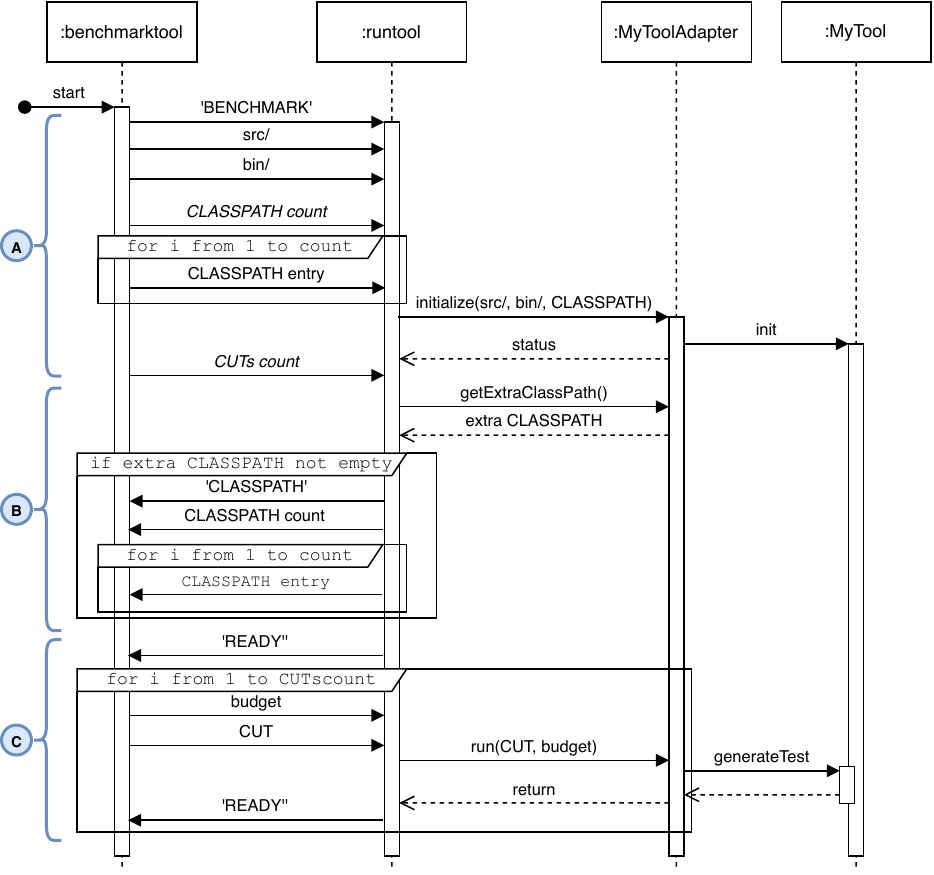}
    \caption{Communication protocol of the \runtool adapter}
    \label{fig:runtool-protocol}
\end{figure}

\paragraph{Communication protocol.}
The adapter has to support the protocol described in \autoref{fig:runtool-protocol}. In the first part (\textit{A}), the \texttt{benchmarktool} signals the start of a new evaluation by sending the '\texttt{BENCHMARK}' message, followed by the paths to the source code and binaries of the software under test, the \texttt{CLASSPATH}, and the number of classes under test in the evaluation. Based on that information, the \runtool adapter initializes the generator (in this case, \mytool).

After the initialization, the generator can signal that it will use additional \texttt{CLASSPATH} entries for its execution. The adapter notifies the \texttt{benchmarktool} of those additional entries (\textit{B} in \autoref{fig:runtool-protocol}). In the third part (\textit{C}), the adapter notifies the \texttt{benchmarktool} that the generator is ready to start the evaluation by sending the \texttt{'READY'} message. The \texttt{benchmarktool} then sends the time budget allocated for the generation and the class under test of the first \textit{run} to the adapter that, in its turn, calls the generator. After the generation, the adapter notifies the \texttt{benchmarktool} that the generator is ready for the next class under test.

%-------------------------------------------
\subsection{Data collection}
\label{subsec:datacollection}
%-------------------------------------------

Once the test cases have been generated, \juge can compute the different \texttt{metrics} for each test suite (\textit{A} in \autoref{fig:architecture}). Those metrics include:
\begin{inparaenum}
\item the number of \emph{flaky and non-compiling} tests, 
\item the \emph{line and branch coverage},
and \item the \emph{mutation score} of the generated tests.
\end{inparaenum}
The \juge infrastructure can be extended to support other kinds of metrics. 

\paragraph{Flaky and non-compiling tests.}
First, if the test suite (one per class under test) does not compile, it is tagged and ignored in the subsequent analysis steps. Once compiled, the test suite is executed five times. Test methods (identified using the \texttt{@Test} annotation) producing different results between different executions are marked as flaky and ignored for the remainder of the analysis. 

\paragraph{Line and branch coverage.}
\juge relies on \jacoco \cite{Hoffmann2014} for statement and conditions coverage of the generated tests. Coverage information is furthermore used to reduce the subsequent mutation analysis time by restricting the execution of the tests against a given mutant to the tests effectively covering the lines modified by the mutant. 

\paragraph{Mutation analysis.}
In the early versions of \juge, we relied on \pitest \cite{Coles2016a} to generate and execute the mutants. However, it raised several issues for unit test generators relying on a dedicated test execution environment. For instance, test cases generated using \evosuite require executing with a dedicated runner to avoid flakiness, handle inputs and outputs, \etc, preventing \juge from using the \pitest environment for test execution. 
To solve this issue, we refined the mutation analysis to use the default test execution environment, supporting ad-hoc test runners. We use \pitest to generate the various mutants and the results of the line coverage to reduce the analysis time by executing only tests reaching the mutated lines against each mutant. Additionally, we set a hard deadline (5 minutes by default) for the mutation analysis to avoid endless executions.

%--------------------------------------------
\subsection{Data analysis}
\label{subsec:dataanalysis}
%--------------------------------------------

The generators can be compared based on quantitative analysis and the different measures collected during the analysis of the generated tests. For that, \juge relies on Friedman's and post-hoc Conover's tests for multiple pairwise comparison \cite{Conover1981} (available as an \texttt{R} script in \juge). The former is a non-parametric test for significance, and it is widely used for multiple-problem analysis, where the problems correspond to the classes under test in our case. A significant $p$-value for this test indicates that the evaluated tools statistically differ \wrt to the overall performance score (\textit{alternative hypothesis}). While Friedman's test does indicate whether the tools in the comparison are statistically different or not, it does not indicate for which pairs of tools such significance holds. Hence, the statistical analysis is complemented by using the post-hoc Conover's test for the pairwise comparison. Notice that the $p$-values produced by the post-hoc test are further adjusted with the Holm-Bonferroni procedure. This procedure corrects the statistical significance level ($p$-value=0.05) in case of multiple comparisons~\cite{Competition2017}.

%-------------------------------------
\subsection{Internal architecture}
\label{subsec:internal}
%-------------------------------------

\begin{figure}[t]
    \centering
    \includegraphics[width=140mm]{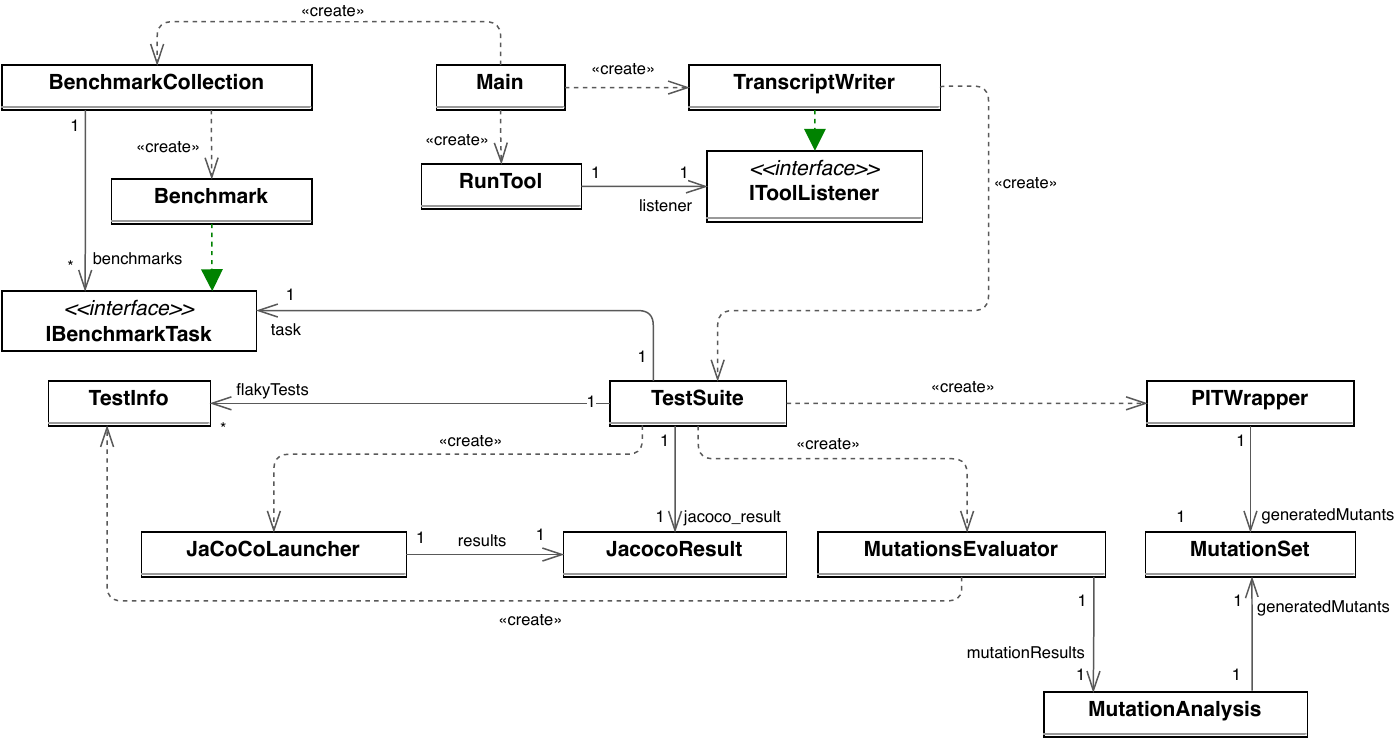}
    \caption{Internal architecture of the \texttt{benchmarktool} module}
    \label{fig:benchmark-class-diagram}
\end{figure}

The module \texttt{benchmarktool} (in Figure \ref{fig:architecture}) is responsible for the orchestration and execution of the different steps of an evaluation. Figure \ref{fig:benchmark-class-diagram} presents the main classes of that module. From a configuration file, the \texttt{Main} class loads the list of the benchmarks (\ie the classes under test) to use for the evaluation in a \texttt{BenchmarkCollection}. Each \texttt{Benchmark} object contains information about the class under test, classpath, and binary and source directory to use. 

Once the benchmarks are loaded, the main class creates a \texttt{RunTool}, which is responsible for handling the execution and communication with the test case generator (the \texttt{sb\-st.ben\-ch\-mark.Run\-Tool} class was renamed to \texttt{:ben\-ch\-mark\-tool} in Figure \ref{fig:runtool-protocol} to avoid confusion with the \texttt{:runtool} adapter module). This class opens an \texttt{SBSTChannel}, implemented as print and write streams on standard input and output, to communicate with the adapter (\texttt{:runtool} in Figure \ref{fig:runtool-protocol}) of the generator following the protocol described in Section \ref{subsec:unittestgeneration}.

For each generated test suite, the \texttt{TranscriptWriter} class is responsible for collecting the different data (encapsulated in \texttt{TestInfo} objects) described in Section \ref{subsec:datacollection}. The \texttt{JaCoCoLauncher} and \texttt{PITWrapper} encapsulate calls to (resp.) \jacoco, for code coverage information, and \pitest, for mutation analysis. As explained in \ref{subsec:datacollection}, it was not possible to use \pitest to execute the full mutation analysis. The \texttt{PITWrapper} is therefore used to create the mutants, while a \texttt{MutationEvaluator} is in charge of executing the test suite on them and collecting the corresponding information.

%%%%%%%%%%%%%%%%%%%%%%%%%%%%%%%%%%%%%%%%%%%%%%%%%%%%%%%%%%%%%%%%%%%%
\section{Setting up and running an evaluation with JUGE}
\label{sec:evaluation}
%%%%%%%%%%%%%%%%%%%%%%%%%%%%%%%%%%%%%%%%%%%%%%%%%%%%%%%%%%%%%%%%%%%%

This section provides general guidelines for evaluating and comparing unit test generation tools with \juge. 

%-----------------------------------------------------
\subsection{Evaluation setup}
\label{subsec:evaluationsetup}
%-----------------------------------------------------

\juge can be used to evaluate automated unit test generators that do not require human intervention during the generation process. It relies on the source code or binaries of a set of projects. \juge is primarily designed to help carry out quantitative studies. It comes with support for structural coverage and mutation analysis and can be extended to support other measures. 

Nevertheless, \juge also allows qualitative analysis as all the tests generated during the evaluation are saved and can be inspected or reused. Additionally, as explained in Section \ref{subsec:unittestgeneration}, \juge allows the unit test generators to save any additional data for later analysis. For instance, a search-based unit test generator can save intermediate fitness values to analyze the fitness landscape evolution. 

\paragraph{Generator meta-parameters.}
Many unit test generators can be configured through meta-parameters (\eg mutation and crossover probabilities for search-based approaches). To ease the evaluation and processing of the results, we recommend considering each configuration as an individual generator with its adapter in a dedicated folder (\textit{A} in \autoref{fig:architecture}) and an explicit name reflecting the configuration. Configuring the generators with the right parameters to answer the research questions and reporting those configurations in the empirical study is of paramount importance to reduce the threats to validity and enable the replicability of the results. 

\paragraph{\juge meta-parameters.}
The infrastructure has two meta-parameters: the \textit{time budget} and the \textit{number of repetitions}. The time budget corresponds to the budget allocated to generate a set of test cases for a given benchmark (\ie a class under test). \juge also uses the time budget to set a global timeout for each execution equal to twice the time budget. The time budget depends mainly on the type of approach used by the test case generator. For instance, previous research indicates that a time budget of three minutes is suited for a search-based generator like \evosuite \cite{Fraser2011, Panichella2018} but is not enough for symbolic execution approaches like \tardis or \sushi \cite{Tool2019SUSHI}.

Similarly, the number of repetitions varies if the generator relies on an exact approach or uses randomness. For exact approaches, one execution is enough (unless one of the research questions considers the execution time, in which case, several repetitions are necessary). For randomized approaches (\eg search-based and random approaches), several repetitions are necessary to ensure the statistical power of the results. Arcuri and Briand \cite{Arcuri2014} estimated that the number of repetitions is a compromise between the number of benchmarks used in the evaluation, the execution time of the generators, and the overall budget available to perform the evaluation. They concluded that each randomized generator should be executed 1,000 times and, if it is not possible, report the reasons and the total execution time of the entire evaluation. However, the number of repetitions (for a more significant number of benchmarks) should be at least 10.

%----------------------------------
\subsection{Benchmarks}
\label{sec:benchmarks}
%----------------------------------

The selection of the benchmarks (\ie sets of classes under test) should follow a systematic approach and ensure that the benchmarks are diverse enough to reduce the threats to the validity of the research questions \cite{Nagappan2013}. For instance, by considering projects from different application domains. 
Those projects (and classes under test) can come from existing benchmarks: \eg \defectsforj \cite{Just2014b} or the previous editions of the tool competition relying on \juge \cite{Competition2016, Competition2017, Competition2018, Competition2019, Competition2020}.

\begin{lstlisting}[float=t,
caption=Excerpt of a \texttt{benchmarks.list} configuration file,
label=lst:benchmarksconfig]
{
  BCEL-1= { (*@\label{line:benchmarksconfig:id}@*)
    src=/var/benchmarks/projects/bcel-6.0-src/src/main/java (*@\label{line:benchmarksconfig:src}@*)
    bin=/var/benchmarks/projects/bcel-6.0-src/target/classes (*@\label{line:benchmarksconfig:bin}@*)
    classes=(org.apache.bcel.classfile.Utility) (*@\label{line:benchmarksconfig:cut}@*)
    classpath=(/var/benchmarks/projects/bcel-6.0-src/target/classes) (*@\label{line:benchmarksconfig:classpath}@*)
  }
  BCEL-2= {
    [...]
  }
}
\end{lstlisting}

The benchmarks are described in a dedicated configuration file (\texttt{benchmarks.list}). \autoref{lst:benchmarksconfig} provides an excerpt of \texttt{benchmarks.list} configuration file from the \juge example benchmarks.
Each benchmark has a unique identifier (line \ref{line:benchmarksconfig:id}), the path to the root folder of the source files of the project (line \ref{line:benchmarksconfig:src}), the path to the root folder containing the compiled classes (line \ref{line:benchmarksconfig:bin}), the list of classes under test (line \ref{line:benchmarksconfig:cut}), and the classpath with all the dependencies to use for the generation and coverage and mutation analysis (line \ref{line:benchmarksconfig:classpath}). 
Once the benchmarks are defined, \juge allows building a new Docker image (\textit{B} in \autoref{fig:architecture}) that can be instantiated multiple times in different containers to run the different tools. 

%---------------------------------------------------------
\subsection{Evaluation execution and results processing}
%---------------------------------------------------------

Once the benchmarks and meta-parameters are defined, \juge can start the evaluation by running different commands from the home directory of the tool in the Docker image (\eg \texttt{/home/mytool} in the example of \autoref{fig:architecture}). We summarize hereafter the main steps and commands to use during the evaluation.\footnote{Details on how to start the Docker container, and the different commands available in \juge are available in the documentation at \url{https://github.com/JUnitContest/JUGE/blob/master/docs/}.} 
If the available hardware allows it, it is possible to run several Docker containers in parallel (instantiated from the same \juge Docker image), each responsible for executing a different generator. One should, however, be cautious to avoid overloading the machine as it could impact the execution of the generators and provoke timeouts. Different Docker containers should be run on independent machines with the same hardware configuration. Practically, if this is not possible, we strongly recommend doing some initial tests to determine the adequate number of parallel Docker containers to avoid undesirable side effects. 

\paragraph{Unit test generation.}
\juge allows running multiple \textit{rounds} of the tool's execution on the same benchmarks and with the same budget in one command. 
Each execution's results are placed in a folder named after the tool and the time budget (\eg \texttt{/home/mytool/results\_mytool\_10} for a time budget of 10 seconds). For each benchmark and each round, \juge creates a folder with the tests generated by the tool (\eg \texttt{BCEL-1\_1}, \texttt{BCEL-2\_1}, \etc for the first round of executions). Those different folders also contain text files with the logs and additional data produced by the generator. 

\paragraph{Data collection and analysis.}
After the generation of the unit tests, \juge can perform a coverage and mutation analysis using \jacoco and \pitest. The different results are stored in a dedicated CSV file (\texttt{transcript.csv}) for analysis. In our future work, we intend to extend \juge to consider other types of measures, for instance, test case readability \cite{Daka2015}. 

Once the different metrics have been computed for the different tools and budgets, the different results can be grouped in a single \texttt{results.csv} for a global quantitative analysis. \juge can perform statistical analysis, as explained in Section \ref{subsec:dataanalysis}, and produce a report with the results of the comparison.

%------------------------------------------------------
\subsection{Reporting, archiving, and reproducibility}
\label{subsec:reporting}
%------------------------------------------------------

One of the goals of the \juge infrastructure is to enhance \emph{repeatability} and \emph{reproducibility} of both the results and statistical and qualitative analysis. For that, we strongly recommend submitting an \emph{artifact} containing the following elements:
\begin{itemize}
    \item the benchmarks, the generated tests, and additional data, if any, 
    \item the files produced by the coverage and mutation, as well as any additional analysis, 
    \item the results of the statistical analysis, together with any other data analysis scripts used for the evaluation.
\end{itemize}
Suppose some of the benchmarks are under a non-disclosure agreement. In that case, we strongly recommend adding benchmarks from open source systems to the evaluation and releasing those in the artifact. The design of such an artifact must be considered early in the study. We recommend, for instance, to fork the \juge repository and update the benchmarks configuration and files to generate a Docker image used to perform the evaluation. The fork can then be easily saved in a data repository (like Zenodo,\footnote{\url{https://zenodo.org}} which has a GitHub integration) for long-term storage with a dedicated DOI. 

In addition to the artifact, the reporting of the evaluation setup should mention the following elements:
\begin{itemize}
    \item the randomized (or not) nature of the generators used in the evaluation;
    \item the meta-parameter configuration(s) of each generator;
    \item the meta-parameter configuration of \juge (including the number of repetitions in the case of randomized generators) with a justification for those values;
    \item the total number of independent executions and the total execution time taken by the evaluation;
    \item the specifications of the hardware and the number of Docker containers running in parallel;
    \item the benchmarks selection procedure and the characteristics of the selected benchmarks relevant to the goals of the evaluation (\eg the number of lines of code of the projects and classes under test, the average McCabe's cyclomatic complexity of the benchmarks, \etc);
    \item any additional data collected and statistical analysis performed on the evaluation results with a proper justification (\eg see Arcuri and Briand \cite{Arcuri2014} for a discussion on statistical analysis for randomized algorithms).
\end{itemize}

%%%%%%%%%%%%%%%%%%%%%%%%%%%%%%%%%%
\section{Impact of JUGE}
\label{sec:impact}
%%%%%%%%%%%%%%%%%%%%%%%%%%%%%%%%%%

The \juge infrastructure played a significant role in the \emph{replication} of previous results regarding the structural coverage and mutation score achieved by automated unit test generators. The configurability of the infrastructure through the meta-parameters and the benchmarks considered for the various editions of the tool competition allowed us to assess the generated tests under various conditions. 
It independently confirmed that
\begin{inparaenum}
\item search-based unit test generation (as implemented in \evosuite) achieves a better coverage and mutation score \cite{Competition2016, Competition2017, Competition2018, Competition2020, Competition2021, Competition2022};
and \item automatically generated tests can compete with manually written ones \wrt coverage and mutation score \cite{Competition2018, Competition2019}. 
\end{inparaenum}

%-----------------------------------------------------------
\subsection{Ten editions of the tool competition}
%-----------------------------------------------------------

The \juge infrastructure and the tool competition also helped to push the boundaries of unit test generation by confronting industrial generators to academic ones and showcasing how research can contribute to the industrial practices \cite{Competition2015, Competition2021, Competition2022}. Also, selecting various benchmarks from open source systems helped to improve the academic generators by confronting them with new classes under test, thereby increasing the generalisability of the underlying approaches. 
For instance, \evosuite has entered the competition multiple times with several algorithms (\textit{whole suite approach} \cite{Fraser2013b}, \mosa \cite{Panichella2015}, \dynamosa \cite{Panichella2018}, \etc) and in 2019, the results of the competition lead to the fix of a major bug \cite{Tool2019EvoSuite}. The results of \evosuite have also been recently independently confirmed using \juge by Herlim \etal \cite{Herlim2021}. 

\begin{table}[t]
    \centering
    \caption{Editions of the tool competitions relying on the \juge infrastructure with the generators, the time budgets (in seconds), the number of classes under test (\#C), and the projects considered for the edition. Industrial tools are indicated by a start ($^{\star}$).}
    \label{tab:competitions}
    \begin{footnotesize}
    \begin{tabularx}{\textwidth}{l >{\raggedright\arraybackslash}X >{\raggedright\arraybackslash}p{12mm} r >{\raggedright\arraybackslash}X}
        \toprule
        \textbf{Edition} 
            & \textbf{Generators} 
            & \textbf{Budgets (in sec.)}
            & \textbf{\#C}
            & \textbf{Projects}
            \\
        \midrule
        2013 \cite{Competition2013} 
            & \randoop, \evosuite \cite{Tool2013EvoSuite}, \ttwo \cite{Tool2013T3} 
            & -
            & 77
            & Apache Commons Lang, Apache Lucene, Barbecue, Joda Time, sqlsheet
            \\
        2014 \cite{Competition2014} 
            & \randoop, \evosuite \cite{Tool2014EvoSuite}, \tthree \cite{Tool2014T3} 
            & -
            & 63
            & Async Http Client, eclipse-cs, GData Java Client, Guava, Hibernate, JMLL, JWPL, Scribe, Twitter4j
            \\
        2015 \cite{Competition2015}  
            & \randoop, \evosuite (whole-suite) \cite{Tool2015EvoSuite}, \evosuite (\mosa) \cite{Tool2015MOSA}, \grt \cite{Tool2015GRT}, \jtexpert \cite{Tool2015JTExpert}, \tthree \cite{Tool2015T3}, undisclosed Commercial Tool (CT)$^{\star}$ 
            & - 
            & 63
            & Async Http Client, eclipse-cs, GData Java Client, Guava, Hibernate, JMLL, JWPL, Scribe, Twitter4j
            \\
        2016 \cite{Competition2016}
            & \randoop, \evosuite (whole-suite) \cite{Tool2016EvoSuite}, \jtexpert \cite{Tool2016JTExpert}, \tthree \cite{Tool2016T3} 
            & 60, 120, 240, 480
            & 68
            & \defectsforj \cite{Just2014b}
            \\
        2017 \cite{Competition2017} 
            & \randoop, \evosuite (whole-suite) \cite{Tool2017EvoSuite}, \jtexpert \cite{Tool2017JTExpert} 
            & 10, 30, 60, 120, 240, 300, 480
            & 69
            & Apache Commons BCEL, Imaging, and Jxpath, Freehep, Gson, Re2J, LA4J, Okhttp 
            \\
        2018 \cite{Competition2018} 
            & \randoop, \evosuite (whole-suite) \cite{Tool2018EvoSuite}, \tthree \cite{Tool2018T3} 
            & 10, 60, 120, 240
            & 59
            & Dubbo, FastJason, JSoup, Okio, Redisson, Webmagic, Zxing 
            \\
        2019 \cite{Competition2019} 
            & \randoop, \evosuite (\dynamosa) \cite{Tool2019EvoSuite}, \sushi \cite{Tool2019SUSHI}, \tardis \cite{Tool2019SUSHI}, \tthree \cite{Tool2019T3} 
            & 10, 60, 120, 240
            & 38
            & Antlr4, AuthzForce, Dub\-bo, Fes\-car, Fa\-st\-Jason, Im\-ixs-Work\-flow, Okio, Spoon, Webmagic, Zxing
            \\
        2020 \cite{Competition2020} 
            & \randoop, \evosuite (\dynamosa) \cite{Tool2020EvoSuite} 
            & 60, 180
            & 70
            & Fescar/Seata, Guava, PdfBox, Spoon
            \\
        2021 \cite{Competition2021} 
            & \randoop, \evosuite \cite{Tool2021EvoSuite}, \evosuitedse \cite{Tool2021EvoSuiteDSE}, \kex~\cite{Tool2021Kex}$^{\star}$, \utbot \cite{Tool2021UtBot}$^{\star}$
            & 30, 120
            & 98
            & Seata, Guava, FastJSON, Spoon, Weka, Okio
            \\
        2022 \cite{Competition2022} 
            & \randoop, \evosuite \cite{Tool2022EvoSuite}, \bbc \cite{Tool2022BBC}, \kex and \kex (reflection) \cite{Tool2022Kex}$^{\star}$, \utbot and \utbot (mocks) \cite{Tool2022UtBot}$^{\star}$
            & 30, 120
            & 65
            & Seata, Guava, FastJSON, Spoon
            \\
        \bottomrule
    \end{tabularx}
    \end{footnotesize}
\end{table}

Table \ref{tab:competitions} describes the main characteristics of the different editions of the tool competition. Over the years, various tools have entered the competition and evolved. Among the different tools, \randoop is used as a baseline, and \evosuite has joined every year since the first edition. In 2015, 2021, and 2022, different industrial test case generation tools entered the competition. 

The different editions have also tried different configurations \wrt to the execution of the tools and the time budget allocated for the generation. Before 2016, the time budget was left to the participants to decide (marked as - in Table \ref{tab:competitions}). Since 2016, the organizers have tried various time budgets to assess how the different tools react under a minimal budget: 10 seconds in 2017 and 2018 and 30 seconds in 2017 and 2021. 

Similarly, the different editions have used classes under tests from various open-source projects to allow the distribution of the benchmarks after the competition. It allows one to replicate the results and the participants of the next edition to try their \runtool adapter before submitting their tool to the competition. In 2016, the organizers used \defectsforj to generate regression tests and assess the tools' capability to expose real-world faults. Also, in 2019, 78 classes under test were initially selected. However, due to issues with metrics computation (that have been fixed), the number of classes considered for the final ranking dropped to 38.

Running the tool competition every year is not trivial. One of the main challenges the different organizers face is the hardware infrastructure required due to the limited time between the submission of the different tools and the limit for providing the results (around two weeks). Both the generation of the tests and their evaluation using coverage and mutation analysis are time-consuming, requiring a powerful server or a cluster.

%---------------------------------------------------------------------
\subsection{Overview of the results of the competition}
%---------------------------------------------------------------------

\begin{figure}[t]
    \centering
    \includegraphics[width=110mm]{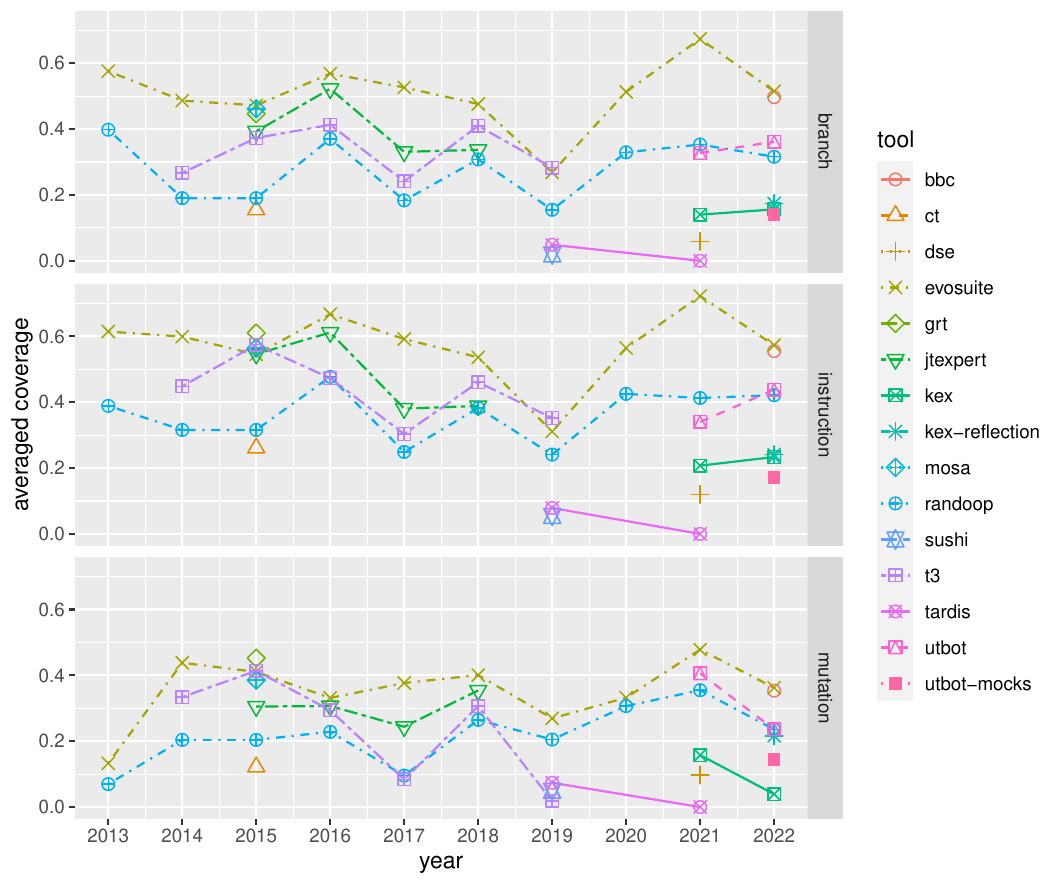}
    \caption{Averaged coverage evolution per tool over the years.}
    \label{fig:avgcoverage}
\end{figure}

As illustrated in Table \ref{tab:competitions}, several tools have entered the competition over the years. Figure \ref{fig:avgcoverage} presents the averaged instruction and branch coverage, and averaged mutation score of the different tools per year, collected from the reports of the past editions \cite{Competition2013, Competition2014, Competition2015, Competition2016, Competition2017, Competition2018, Competition2019, Competition2020, Competition2021, Competition2022, Tool2013EvoSuite, Tool2013T3, Tool2014EvoSuite, Tool2014T3, Tool2015EvoSuite, Tool2015GRT, Tool2015JTExpert, Tool2015MOSA, Tool2015T3, Tool2016EvoSuite, Tool2016JTExpert, Tool2016T3, Tool2017EvoSuite, Tool2017JTExpert,  Tool2018EvoSuite, Tool2018T3, Tool2019EvoSuite, Tool2019SUSHI, Tool2019T3, Tool2020EvoSuite, Tool2021EvoSuite, Tool2021EvoSuiteDSE, Tool2021Kex, Tool2021UtBot, Tool2022BBC, Tool2022EvoSuite, Tool2022Kex, Tool2022UtBot}. As can be seen from the Figure, \evosuite has the best averaged structural coverage and mutation scores over the years. The evolution from one edition to another can be attributed to several factors, including bug correction and improvements of the underlying algorithms, selection of different time budgets and benchmarks. The full dataset providing averaged coverage, mutation score, and execution times per benchmark and time budget is available on Zenodo \cite{xavier_devroey_2022_7194656}.\footnote{Due to incomplete data, the results of T2 from 2013\cite{Tool2013T3} are not included in Figure \ref{fig:avgcoverage}.}

\begin{figure}[t]
    \centering
    \includegraphics[width=110mm]{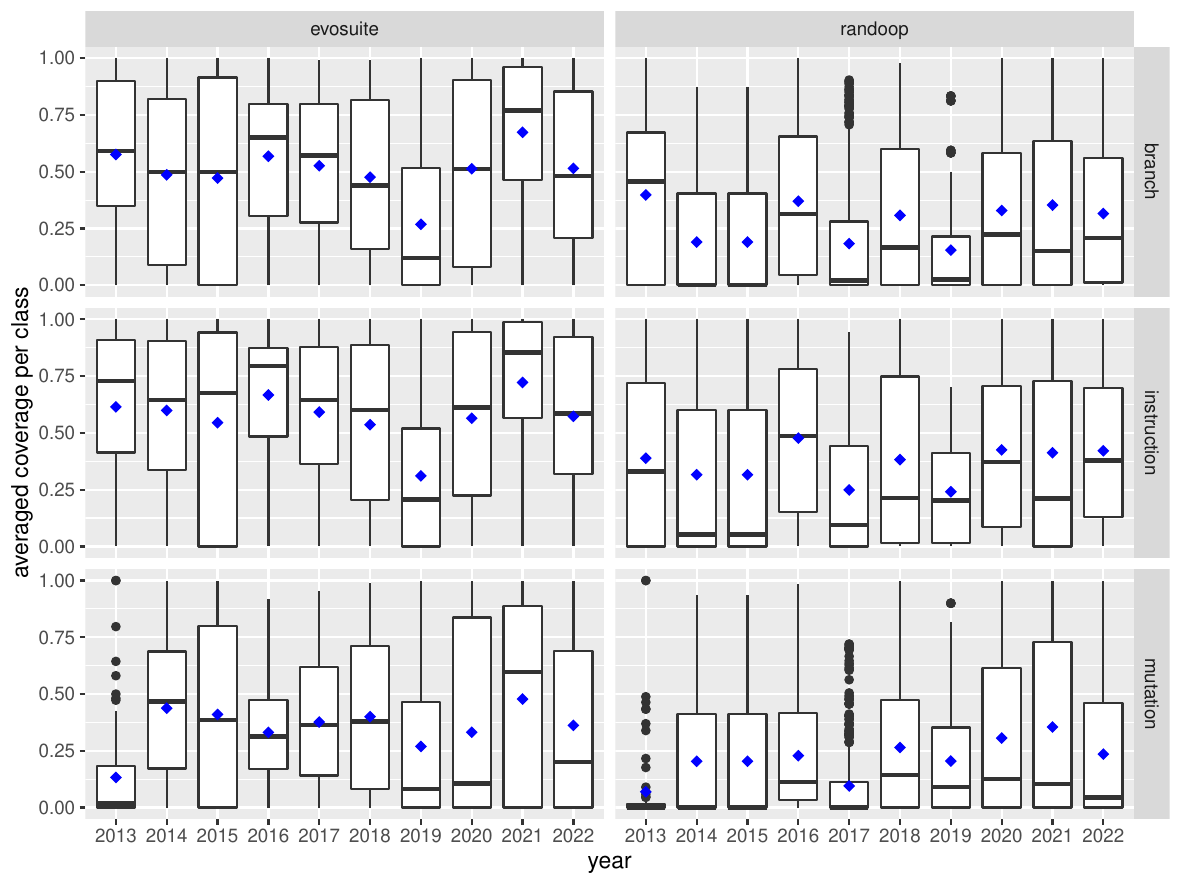}
    \caption{Averaged coverage evolution per benchmark for EvoSuite and Randoop. Diamond (\textcolor{blue}{$\blacklozenge$}) indicates the mean value.}
    \label{fig:avgcoverageevosuiterandoop}
\end{figure}

Among the different tools, \randoop, relying on feedback-directed random test generation \cite{Pacheco2007}, and \evosuite, relying on genetic algorithms \cite{Fraser2011}, have entered the competition every year since 2013. Figure \ref{fig:avgcoverageevosuiterandoop} presents the distributions of the averaged coverage evolution per benchmark for each configuration of \evosuite and \randoop for the different editions of the competition (\ie each data point represents an averaged value of all the executions of the tool with one time budget on one benchmark). Generally, \evosuite performs better on individual benchmarks than \randoop for unit test generation. This is confirmed when applying Friedman's non-parametric with post-hoc Nemenyi test~\cite{Japkowicz2011,Panichella2021} on instructions coverage (\randoop ranking 1.77 and \evosuite 1.23), branch coverage (\randoop ranking 1.77 and \evosuite 1.23), and mutation score (\randoop ranking 1.73 and \evosuite 1.27). The rankings are significantly different as their corresponding average ranks differ by at least the given critical distance, here, 0.048, with a Friedman's test p-value below 0.01 for the three cases.

%%%%%%%%%%%%%%%%%%%%%%%%%%%%%%%%%%%%%%%%%%%%%
\section{Discussion and lessons learned}
\label{sec:discussion}
%%%%%%%%%%%%%%%%%%%%%%%%%%%%%%%%%%%%%%%%%%%%%

Any empirical evaluation of automated unit test generation faces several technical and methodological challenges. \juge seeks to alleviate those challenges by providing a standardized way of designing, running, and reporting such evaluations. 
Both the development of \juge and the evaluation method reported in Section \ref{sec:evaluation} took several years to develop. We discuss hereafter the main lessons learned and potential new applications of \juge. 

% ---------------------------
\subsection{Building \juge and evaluating generators}
% ---------------------------

\paragraph{Adapt to different generators.}
The main technical challenges for such an infrastructure come from the diversity of the generators that can be considered for an empirical evaluation (\ie random-based, search-based, concolic/symbolic-based, \etc)\challenegeextensible. It requires \textit{isolating} \challenegeisolation the executions to avoid troubles in case of a bug in the generator (\eg erasing files from the host file system \cite{Fraser2013}) while still having a \textit{standard communication interface} \challenegestandard. It is achieved through an adapter with a shared standard set of commands used by \juge to interact with the generator. \juge runs in a Docker container to guarantee the isolation of the generator from its environment during test case generation. 

\paragraph{Performance, scalability, and statistical power.}
As for any empirical evaluation with a random-based generator, researchers have to balance the number of classes under test to consider reducing external validity with the number of executions to ensure enough statistical power, giving the external constraints on the overall execution time \cite{Arcuri2014}. For instance, in the tool competition, the entire evaluation must be done in around two weeks. To cope with this limitation, organizers use sampling to select a subset of classes under test, limit the time budget (not more than 8 minutes), and the number of repetitions of the executions (between 6 and 10, depending on the year). As explained in Section \ref{subsec:evaluationsetup}, the time budget allocated to the generator depends on the type of approach and the goals of the evaluation. 

\juge supports parallel executions by running several Docker containers in parallel (the number of containers depends on the technical specifications of the host machine), allowing to \textit{scale} while preserving the overall \textit{performances} of an evaluation \challenegeperformance. It allows relying on standard Docker technologies and tools (\eg dashboards) to handle the overall evaluation. 

\paragraph{Configuration of the meta-parameters.}
In addition to the time budget and the number of repetitions, which can be configured for \juge, the generators usually come with various meta-parameters that directly influence the generation process \challenegeextensible. As explained in Section \ref{subsec:evaluationsetup}, such parameters should be carefully considered and reported to reduce the threats to the validity and enable the \textit{replicability} of the results. 

For instance, many test generators like \evosuite and \randoop include post-processing mechanisms that can be activated to minimize the generated tests \cite{Fraser2011, Pacheco2007}. Such mechanisms are time-consuming and can be deactivated to reduce the overall execution time when evaluating properties such as coverage or the mutation score. However, deactivating test case minimization has a significant impact on other properties, such as structural properties, readability, the execution time of the tests, \etc Researchers should be aware of such impacts and carefully consider them when designing their studies. 

\paragraph{Analysis of the generated tests.}
Automated test case generation is a challenging task and requires several mechanisms (\eg code instrumentation, handling I/O operations on the system under test, \etc) to be effective. Among the possible mechanisms is using a specific scaffolding for the generated tests: for instance, \evosuite controls elements that could be non-deterministic to avoid test flakiness. 
However, such mechanisms might cause undesired interactions with the infrastructure and, more specifically, with the tools used to analyze the generated tests \challenegeextensible. It has been the case for \evosuite and the mutation analysis: the test runner (\texttt{EvoRunner}) used in the generated tests was not compatible with \pitest and required to use the mutated \texttt{.class} files directly instead of relying on the optimized \pitest infrastructure. In the latest version, \juge includes options to parallelize the execution of the mutation analysis and reduce the overall execution time of the evaluation \challenegeperformance. 

\paragraph{Repeatability and reproducibility.}
\juge considers each generator configuration (\eg \evosuite using a different generation algorithm) as a generator that will require its \runtool adapter and corresponding shared folder, like, for instance, the different algorithms used with \evosuite in the competition. Those generators should be shared in a companion artifact to provide \textit{standard} readily usable implementations to the research community \challenegestandard. As one of the main goals of \juge is to provide a common platform where researchers and practitioners can plug their generators and compare them using various benchmarks, sharing generators and their corresponding adapters will greatly improve \textit{repeatability} (\ie the same evaluation can be performed by the same team in the same conditions and produce the same results) and \textit{reproducibility} (\ie the same evaluation can be performed by a different same team in similar conditions and produce the same results) of the results. 

% ----------------------------------
\subsection{Benchmarks selection}
\label{subsec:benchmarkselection}
% ----------------------------------

As explained in Section \ref{sec:benchmarks}, \juge allows one to define their own set of benchmarks for a given evaluation. Section \ref{subsec:benchmarksselection} provides an example of benchmarks selection process, followed in the eighth edition of the tool competition \cite{Competition2020}. As illustrated by the example, the competition mainly used an opportunistic approach by considering open-source projects built with Maven to ease the collection of projects' dependencies. The main selection criteria were availability and the complexity of the classes to test. Other criteria can also be taken into account.

In general, selecting representative benchmarks for an evaluation is not an easy task. The selection criteria depend on the goals of the evaluation but also the availability of the benchmarks. We provide some guidelines for benchmark selection based on our experience and related literature.
    
\paragraph{Representative.} 
A good set of benchmarks should be representative of real-world software. This is a best practice adopted by the software engineering research community to ensure that the evaluation addresses a relevant problem \cite{Dyba2005, Metzman2021, Vos2012, Juristo2004a, Panichella2018}. The benchmarks can come from one or more systems depending on the experimental design. For instance, Almasi \etal \cite{Almasi2017} performed an evaluation of unit test generation on a closed-source industrial \textit{case study}. Other evaluations studied unit test generation on various open-source (and openly accessible) systems \cite{Panichella2015, Panichella2018, Fraser2013b, Campos2018}. In this latter case, one should take care of selecting systems that are popular (\eg by looking at the number of stars on GitHub or the number of dependant projects in Maven central) and under active maintenance (\eg by looking at the commit frequency) to ensure that the results of the evaluation will be relevant to the software engineering industrial community. 

\paragraph{Diverse.}
The benchmarks used for an evaluation should be diverse enough to mitigate threats to external validity. This diversity can, for instance, be achieved by considering benchmarks from multiple projects from diverse application domains. Another approach could be, instead of a random sampling like in Section \ref{subsec:benchmarksselection}, selecting diverse benchmarks based on the coverage and mutation score a random generator achieves.

\paragraph{Adequate.} 
The benchmarks should be selected adequately \wrt the goals of the evaluation. For instance, if one of the evaluation goals is to compare the coverage and mutation score of automatically generated and hand-written tests, one has to consider the coverage and mutation score of the existing test suites when selecting the benchmarks. \Eg in 2019, the competition \cite{Competition2019} introduced Spoon in the benchmarks as it has a hand-written test suite with high coverage. The results showed that none of the generators used in that edition could achieve a higher branch or line coverage, or mutation score. 

\paragraph{Challenging.} 
Depending on the kind of generation technique considered, some benchmarks might not be relevant or do not bring any interesting insights. For instance, Shamshiri \etal \cite{Shamshiri2015} showed that many classes in the SF110 corpus \cite{Fraser2014b}, a corpus of benchmarks widely used for unit test generation, can be covered using random search. Those benchmarks should be filtered out for evaluating search-based unit test generators as they will not provide interesting insights. For instance, the competition and other related work \cite{Panichella2018, Campos2018} have considered McCabe's cyclomatic complexity (\ie the number of independent paths in a control flow graph) of the benchmarks to filter out simple benchmarks, easily covered (\ie classes with a cyclomatic complexity lower than five).

\paragraph{Comparable.}
Generalizability is hard to achieve and requires several evaluations, preferably performed independently by different research teams. Selecting a subset of benchmarks that have already been used in other studies is interesting to allow comparison between different studies. It enables the meta-analysis of the results and cross-comparisons between different studies. For instance, Campos \etal \cite{Campos2018}, and Panichella \etal \cite{Panichella2018} reused the same benchmarks to perform large-scale empirical evaluations of search-based unit test generation algorithms, allowing a direct comparison and discussion of the results.

\paragraph{Documented.}
As stated in Section \ref{subsec:reporting}, both the selection procedure and the characteristics of the benchmarks should be reported together with the evaluation results. It is essential to the reviewing process to ensure that reviewers have enough information about the benchmarks to assess the paper describing the evaluation \cite{Ralph2020a}, but also to ensure that future research can compare and replicate the results on other benchmarks with similar characteristics. 

\paragraph{Available.}
Finally, benchmarks coming from open-source projects should be made available to the research community. Building a good benchmark is not trivial and represents a substantial effort \cite{Just2014b}. The community should share this effort by encouraging the best open-science practices and distribution of the benchmarks. \juge provides a standard way of sharing and reusing such benchmarks (described in Section \ref{sec:benchmarks}). \footnote{Benchmarks of previous tool competitions are available at \url{https://github.com/JUnitContest/JUGE/tree/master/infrastructure}.} 

% ----------------------------------
\subsection{Future applications}
% ----------------------------------

The method described in Section \ref{sec:evaluation} constitutes a standard that can be applied to unit test generation for other kinds of languages using an infrastructure similar to \juge. For instance, Lukasczyk \etal \cite{Lukasczyk2020} recently defined an approach to generate unit tests for Python. Of course, dynamically typed languages such as Python face additional challenges than those discussed in Section \ref{subsec:requirements}. Such challenges must be considered in the design of the infrastructure (\eg running type inference engines during pre-processing) and the selection of the benchmark (\eg considering classes with type annotations only, \etc), and reported in the description of the empirical evaluation.

Besides comparing unit test generators, the \juge infrastructure can be used to generate large amounts of tests for various kinds of classes using different tools and configurations. It enables the continuous creation of an openly available corpus of automatically generated unit tests. Such a corpus would 
\begin{inparaenum}[(i)]
\item directly contribute to the body of empirical evidence on which decision-makers can rely to assess the usage of a unit test generator in their industrial context \cite{Dyba2005};
and \item enables further empirical evaluations on automatically generated tests without configuring and running the generators, which require a certain level of expertise. 
\end{inparaenum}
For instance, in a recent study, Panichella \etal \cite{Panichella2020} revisited previous studies on the presence of test smells in automatically generated tests and found that previous results vastly overestimated their presence. Among the different problems, they pointed out a misconfiguration of \evosuite and its minimization process, resulting in more prominent test cases more likely to contain certain smells. 
Building openly available corpora using the appropriate configuration for the generators, with a description of the characteristics applicable evaluations, would avoid such issues. 
More in general, having the JUGE infrastructure and its associated standards available to industrial and academic research communities can open the road for more systematic testing for other languages (e.g., Python, etc.) as well as the definition of testing pipelines to identify bugs and imperfections of systems in other application domains \cite{UAV:2022,ZAMPETTI2022111425,3533818,Birchler2022Cost}. The availability of such technologies can also impact computer science education, with available tools that can be used in practical courses. Finally, we also expect future investigation supported by JUGE in the context of test code quality and cost-effectiveness of both automatically and manually generated tests \cite{emse:smells-2022,8865437}.

Finally, regarding the infrastructure itself, we plan to add other types of analysis in addition to coverage and mutation score. Such analyses include performances of the generated tests (\eg execution time and memory consumption) and readability.

%%%%%%%%%%%%%%%%%%%%%%%%%%%%%%%%%%
\section{Conclusion}
\label{sec:conclusion}
%%%%%%%%%%%%%%%%%%%%%%%%%%%%%%%%%%

\juge sets a standard for properly assessing automated test case generators. It provides an infrastructure and a method to design, set up, and execute an empirical evaluation, collect and analyze the results, and produce a replication package to meet the requirements, enabling an effective contribution to the software testing empirical body of knowledge. 
It includes recommendations for selecting benchmarks and the parametrization of the generator and the infrastructure, depending on the research questions.
\juge was initially introduced and developed in the context of the tool competition and has been used with several generators and dozens of classes under test from various projects. 

Finally, the \juge infrastructure availability opens several directions for practitioners, who can rely on a large body of empirical evidence to assess automated test case generation usage in their context, and researchers, who can benefit from corpora of automatically generated tests for further empirical evaluations. \juge also provides guidelines for evaluating unit test generation in other programming languages and for defining similar infrastructures in other domains.

%%%%%%%%%%%%%%%%%%%%%%%%%%%%%%%%%%%%%%%%%%%%%%%%%%%%%%%%%%%%%%%%%%%%%%%%%%%%%

\ack We would like to thank (in alphabetical order) Arthur Baars, Sebastian Bauersfeld, Matteo Biagiola, Ignacio Lebrero, Urko Rueda Molina, and Fiorella Zampetti for their contribution to the implementation of the \juge infrastructure. 
We would also like to thank (in alphabetical order) Azat Abdullin, Marat Akhin, Giuliano Antoniol, Andrea Arcuri, Cyrille Artho, Mikhail Belyaev, Pietro Braione, Nikolay Bukharev, José Campos, Nelly Condori, Christoph Csallner, Giovanni Denaro, Gordon Fraser, Yann-Gaël Guéhéneuc, Masami Hagiya, Mainul Islam, Dmitry Ivanov, Gunel Jahangirova, Kiran Lakhotia, Ignacio Manuel Lebrero Rial, Lei Ma, Alexey Menshutin, Arsen Nagdalian,  Gilles Pesant, Simon Poulding, Wishnu Prasetya, Vincenzo Riccio, José Miguel Rojas, Abdelilah Sakti, Hiroyuki Sato, Sebastian Schweikl, Gleb Stromov, Yoshinori Tanabe, Paolo Tonella, Artem Ustinov, Sebastian Vogl, Tanja  Vos, Mitsuharu Yamamoto, Fiorella Zampetti, and Cheng Zhang for their participation in previous editions of the competition and the feedback they provided on the infrastructure. 
Xavier Devroey was partially funded by the EU Horizon 2020 ICT-10-2016-RIA ``STAMP'' project (No.731529), the Vici ``TestShift'' project (No. VI.C.182.032) from the Dutch Science Foundation NWO, and the CyberExcellence (No. 2110186) project, funded by the Public Service of Wallonia (SPW Recherche).
Alessio Gambi's work was partially supported by the DFG project STUNT (DFG Grant Agreement n. FR 2955/4-1).
Sebastiano Panichella and Annibale Panichella gratefully acknowledge the Horizon 2020 (EU Commission) support for the project \textit{COSMOS} (DevOps for Complex Cyber-physical Systems), Project No. 957254-COSMOS.
Ren{\'e} Just's work is partially supported by the National Science Foundation under grant CNS-1823172.

%%%%%%%%%%%%%%%%%%%%%%%%%%%%%%%%%%%%%%%%%%%%%%%%%%%%%%%%%%%%%%%%%%%%%%%%%%%%%

\bibliographystyle{wileyj} 
\bibliography{references}

%%%%%%%%%%%%%%%%%%%%%%%%%%%%%%%%%%%%%%%%%%%%%%%%%%%%%%%%%%%%%%%%%%%%%%%%%%%%%

\appendix

%%%%%%%%%%%%%%%%%%%%%%%%%%%%%%%%%%
\section{Example of Evaluation using JUGE}
\label{sec:example}
%%%%%%%%%%%%%%%%%%%%%%%%%%%%%%%%%%

This appendix illustrates how \juge can be used in practice. We use as an example the eighth edition of the Java unit testing tool competition \cite{Competition2020} in which three of the authors were involved. The competition has occurred since 2013 and is co-located with the Search-Based Software Testing workshop. Participants willing to enter the competition have to provide
\begin{inparaenum}
    \item an executable version of their tool (potentially obfuscated, which enables the participation of industrial tools); 
    and \item an implementation of the \texttt{runtool} adapter (described in Section \ref{subsec:unittestgeneration}) for their tool.
\end{inparaenum}
In practice, this implementation consists in cloning the \texttt{runtool} project available in the \juge repository and defining the methods of the \texttt{ITestingTool} interface described in Listing \ref{lst:ITestingTool}. The effort is minimal and consists of writing between 100 and 200 lines of code for a reasonably well-implemented tool (\ie tools relying on configuration parameters, for instance).

The competition aims to compare different tools on a diverse set of classes under test (\ie benchmarks) for different time budgets. As the organizers run the evaluation, the time budgets and number of executions depend on the computational resources available. For the eighth edition, the tools were executed ten times for 60 and 180 seconds on each benchmark. 

%-------------------------------------
\subsection{Benchmarks selection}
\label{subsec:benchmarksselection}
%-------------------------------------

Several ways exist to define a new set of benchmarks for a given set of projects. The projects selected for that edition were a mix of projects from the previous iterations of the competition to allow comparing results across years and popular GitHub projects built using Maven to ease dependency collection. The benchmarks were selected following a two-step procedure for the list of selected projects. In the first step, 
\begin{inparaenum}
\item identifying the packages in the project that contain classes relevant for the evaluation (\eg packages containing classes with the business logic);
\item computing the McCabe's cyclomatic complexity for the different classes of those packages and remove classes with a complexity lower than five. This reduces the risk to sample classes with few branches, easily covered by randomly generated tests \cite{Panichella2018}.
\end{inparaenum}

\begin{figure}
    \centering
    \includegraphics[width=110mm]{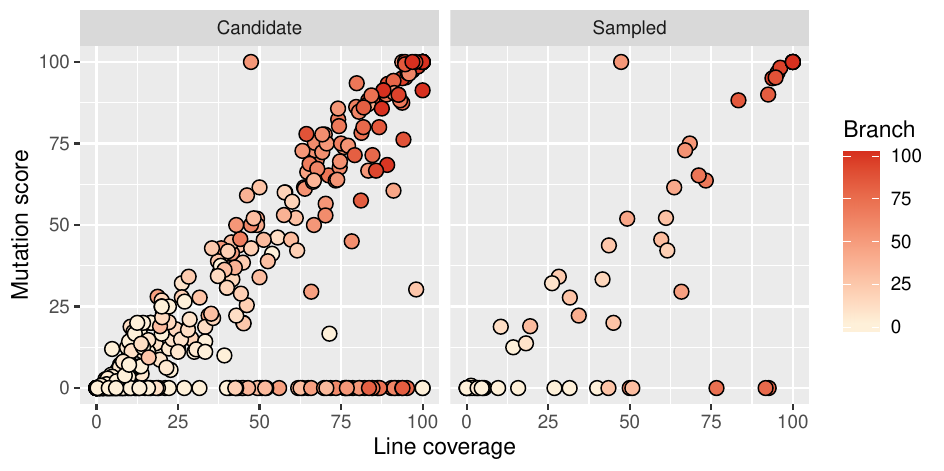}
    \caption{Example of reporting of the line coverage, branch coverage and mutation score for the candidate (382 classes) and selected benchmarks (60 classes, 20 per project) from the 2020 tool competition \cite{Competition2020}.}
    \label{fig:example-cuts}
\end{figure}

In the second step, a random generator (here, \randoop \cite{Pacheco2007}) was executed with a low time budget (\eg ten seconds) on the remaining classes to filter out classes for which the generator could not generate any tests. This reduces the chances of facing technical difficulties while evaluating the different tools. Since the remaining classes were still too high, a subset of classes was randomly sampled for each project. 

In addition to the previous steps, one can also use \juge to perform a coverage and mutation analysis of the tests produced by the random generator and report the results for the candidate and sampled classes. Figure \ref{fig:example-cuts} shows the line and branch coverage and mutation score of the candidate and sampled classes of the eighth edition of the tool competition.

%-------------------------------------
\subsection{Execution}
%-------------------------------------

The eighth edition of the competition received only one submission (\ie \evosuite with \dynamosa) that was compared against \randoop, used as a baseline. The implementation of the \texttt{runtool} adapter is available in \juge's GitHub repository. \evosuite and \randoop were executed ten times each against each benchmark. The executions were run in parallel (using Docker) on two servers: one with 40 CPU cores (2.30GHz) with 128 GB memory and one with 8 CPU cores (2.49GHz) with 160 GB memory. The coverage and mutation analysis were performed on the same machines. The total execution time for test generation and analysis took around four days.

%-----------------------------------------------------------
\subsection{Data analysis and ranking of the contestants}
%-----------------------------------------------------------

\begin{table}[t]
    \centering
    \caption{Example of scores and ranking obtained through Friedman's test for the 5th edition of the tool competition \cite{Competition2017}.}
    \label{tab:examplescore}
    \begin{tabular}{c c c c}
        \toprule
        \textbf{Tool} & \textbf{Score} & \textbf{Std.dev} &\textbf{Ranking} \\
        \hline
        \evosuite   & 1457  & 192.72    & 1.55 \\
        \jtexpert   & 849   & 102.03    & 2.71 \\
        \tthree     & 526   & 82.43    & 2.81 \\
        \randoop    & 448   & 34.75    & 2.92 \\
        \bottomrule
    \end{tabular}
\end{table}

\begin{table}[t]
    \centering
    \caption{Example of results of the post-hoc Conover's test for the 5th edition of the tool competition \cite{Competition2017}.}
    \label{tab:exampleconover}
    \begin{tabular}{c c c c c}
        \toprule
             & \evosuite & \jtexpert & \tthree & \randoop \\
        \hline
        \evosuite   & -         & -         & -         & - \\
        \jtexpert   & $<0.01$   & -         & -         & - \\
        \tthree     & $<0.01$   & $0.01$    & -         & - \\
        \randoop    & $<0.01$   & $<0.01$   & $0.06$    & - \\
        \bottomrule
    \end{tabular}
\end{table}

The competition combines the different metrics to ease the comparison of different generators. For that, it relies on a \emph{scoring formula} \cite{Competition2017}. This formula has been developed and refined during the different editions of the competition and takes into account the line and branch coverage, the mutation score, and the time budget used by the generator, and applies a penalty for flaky and non-compiling tests. In 2020, \evosuite ranked first.
As a further example, Table \ref{tab:examplescore} provides the ranking obtained through Friedman's test for the fifth edition of the tool competition \cite{Competition2017}. \evosuite is ranked first with an average score of 1457, followed by \jtexpert, \tthree, and \randoop. Table \ref{tab:exampleconover} gives the post-hoc Conover's test results for the same edition of the competition and indicates that the various comparisons are statistically significant, except for \tthree and \randoop for which the $p$-value is above the confidence level of $0.05$.

The scoring formula used in the competition provides an aggregated measure to rank the different tools based on their coverage and mutation analysis performances. Other aspects could be considered, like, for instance, the readability of the generated tests. Considering such aspects and including them in the scoring formula is part of the future work identified in the tenth edition of the competition \cite{Competition2022}. 

Once the coverage and mutation analysis have been performed, the full results for a tool (\ie benchmarks, generated tests, and collected data) are sent to the tool's authors for further analysis. Authors can further analyze the results for each round of execution of their tool on each benchmark.

\end{document}